# Multiscale Metrology and Optimization of Ultra-Scaled InAs Quantum Well FETs

*Neerav Kharche[1,2], Gerhard Klimeck[1], Dae-Hyun Kim[3,4], Jesús. A. del Alamo[3], and Mathieu Luisier[1]*

[1] Network for Computational Nanotechnology, Birck Nanotechnology Center, Purdue University, West Lafayette, IN 47907;
[2] Computational Center for Nanotechnology Innovations, Rensselaer Polytechnic Institute, Troy, NY 12180;
[3] Microsystems Technology Labs, Massachusetts Institute of Technology, Cambridge, MA 02139;
[4] Teledyne Scientific & Imaging, LLC, Thousand Oaks, CA, 91360;

*Abstract*—A simulation methodology for ultra-scaled InAs quantum well field effect transistors (QWFETs) is presented and used to provide design guidelines and a path to improve device performance. A multiscale modeling approach is adopted, where strain is computed in an atomistic valence-force-field method, an atomistic $sp^3d^5s^*$ tight-binding model is used to compute channel effective masses, and a 2-D real-space effective mass based ballistic quantum transport model is employed to simulate three terminal current-voltage characteristics including gate leakage. The simulation methodology is first benchmarked against experimental *I-V* data obtained from devices with gate lengths ranging from 30 to 50 nm. A good quantitative match is obtained. The calibrated simulation methodology is subsequently applied to optimize the design of a 20 nm gate length device. Two critical parameters have been identified to control the gate leakage magnitude of the QWFETs, (i) the geometry of the gate contact (curved or square) and (ii) the gate metal work function. In addition to pushing the threshold voltage towards an enhancement mode operation, a higher gate metal work function can help suppress the gate leakage and allow for much aggressive insulator scaling.

## I. Introduction

As Si CMOS technology approaches the end of the ITRS roadmap, the semiconductor industry faces a formidable challenge to continue transistor scaling according to Moore's law [1]. Several industry and academic research groups have recently demonstrated high mobility III-V quantum well field-effect transistors (QWFETs), which can achieve high-speed operation at low supply voltage for applications beyond the reach of Si CMOS technology [2-8]. In particular, InGaAs and InAs channel QWFETs scaled down to 30 nm gate lengths have been shown to exhibit superior performance than Si MOSFETs and their heterogeneous integration on Si substrate has already been demonstrated [3-8].

Device simulations provide useful insights into the operation of QWFETs and might guide experimentalists in the process of scaling their gate length below 30 nm [9, 10]. In this paper, the performance of Schottky gated InAs QWFETs is analyzed using quantum mechanical simulations.

Classical approximations such as the drift-diffusion model can neither capture the quantization of the energy levels resulting from the strong confinement of the electrons in a quantum well nor the tunneling currents in nano-scale devices. To address these limitations quantum mechanical approaches based on the effective mass approximation [10] and on the tight-binding method [9] have already been proposed. While both approaches agree well with experimental data above threshold, they are not able to reproduce the OFF-current region where gate leakage currents dominate. The absence of a real dielectric layer between the channel and the gate contact, contrary to MOSFETs, makes the III-V QWFETs very sensitive to gate leakage currents. To properly account for this effect, a multi-port, two-dimensional (2-D) real-space Schrödinger-Poisson solver based on the effective mass approximation [11] has been developed. Band-to-band tunneling leakage and impact ionization do not have significant effect on IVs of the QWFETs in the operating range considered in this work and they are not included in the transport model. Hence, the OFF- and gate leakage currents are equivalent.

This paper is an expanded version of a recent conference proceeding [11]. A detailed discussion of simulation methodologies and mechanisms behind reported scaling trends in [11] are provided. The paper is organized as follows: in Section II, the device structure and its analysis through a decomposition into intrinsic and extrinsic simulation domains are introduced. Section III describes the core techniques used in this work: the 2-D real-space Schrödinger-Poisson solver based on the effective mass approximation, the tight-binding technique used to calculate the channel effective masses [12-14], and the Newton scheme employed to calibrate the simulator to experimental data. In Section IV, the simulator is benchmarked against the measured characteristics of InAs QWFETs with gate lengths ranging from 30 to 50 nm [7]. The calibrated simulator is subsequently used to optimize the logic performance of an InAs QWFET with a gate length scaled to 20 nm. Finally, the conclusions and outlook of this work are presented in Section V.

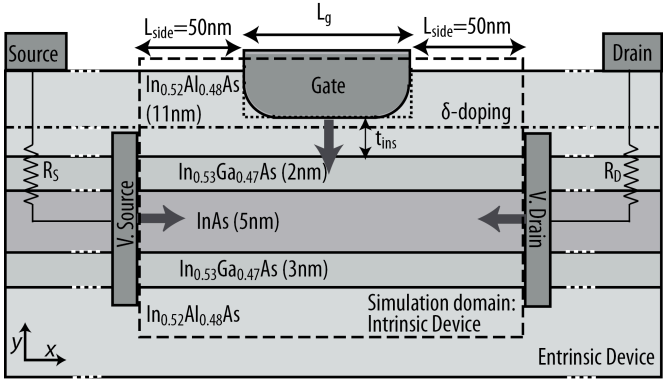

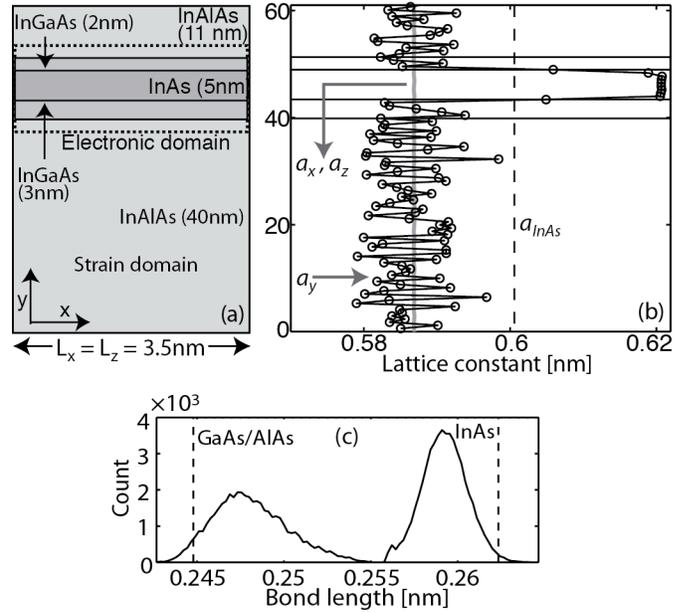

Fig. 1. Schematic view of an InAs QWFET. The dashed black rectangle encloses the quantum transport simulation domain i.e. the intrinsic device. Thick black arrows depict the direction of the electron injection from the virtual contacts into the simulation domain. The source/drain extensions beyond the virtual contacts are modeled by two series resistances $R_S$ and $R_D$, respectively. Two gate contact geometries curved (solid line) and flat (dotted line) are investigated.

Fig. 2. Strain modeling of the QWFET channel. (a) Strain simulation domain to compute the relaxed atom positions in the InAs quantum well channel. The electronic structure calculation domain is depicted by a dotted rectangle..(b) In-plane lattice constant ($a_x$, $a_z$) through the center of the InAs quantum well and lattice constant along the growth direction ($a_y$) through the center of the heterostructure in (a). The unstrained lattice constant of InAs is labeled as $a_{InAs}$. (c) Bi-modal bond length distribution in the InAlAs and InGaAs layers. The bond lengths of pure Ga-As/Al-As and In-As are shown by the dotted lines. Note that the Ga-As and Al-As bond lengths are almost the same.

## II. DEVICE DESCRIPTION

The InAs QWFET [7] considered in this work is schematically shown in Fig. 1. The channel region is composed of a 10 nm $In_{0.53}Ga_{0.47}As/InAs/In_{0.53}Ga_{0.47}As$ (2/5/3 nm, from top to bottom) quantum-well grown on a 500 nm thick $In_{0.52}Al_{0.48}As$ layer lattice matched to InP. The $In_{0.52}Al_{0.48}As$ layer between the quantum-well channel and the gate contact acts as an insulator or barrier. A Si δ-doped layer of concentration $3 \times 10^{12}$ cm$^{-2}$ situated 0.3 nm below the gate contact supplies the channel conduction electrons. The source/drain contacts are located on the top of the device almost 1 μm away from the gated region.

To reduce the computational burden, this structure is analyzed by breaking it into two distinct domains. The intrinsic simulation domain is restricted to under the gate contact and an extension $L_{side}$ of 50 nm on each side. The ideal contacts, labeled as the virtual source/drain in Fig. 1, are placed at the two ends of the intrinsic device. The part of device outside the intrinsic domain is labeled as the extrinsic domain, which is modeled via two series resistances $R_S$ and $R_D$ following the procedure described in Ref. [15]. Due to the idealized contact assumption, phenomena such as "source starvation" are not included in our simulations [16].

Two gate contact geometries, curved and flat, resulting from different gate-stack fabrication processes are considered. The edges of the curved gate contact are quarter circles with the radius of curvature equal to $t_{InAlAs}-t_{ins}$, where, $t_{InAlAs}$ is the total thickness of the top InAlAs layer and $t_{ins}$ is the thickness of the InAlAs layer between the gate contact and the quantum-well channel. Experimentally, a curved gate contact geometry is expected from the isotropic wet chemical etching process that is used to recess the gate [7].

## III. APPROACH

A three-step multiscale modeling approach is adopted to simulate the InAs QW FETs. First the strain arising from the growth of an InAs layer in the middle of two $In_{0.53}Ga_{0.47}As$ layers is computed by an atomistic valence-force-field (VFF) method. Then an atomistic tight-binding method is used to calculate the electron dispersion in the quantum well channel and the corresponding electron effective masses. In a third step, these effective masses are inserted into an effective mass based quantum transport simulator that yields the current-voltage characteristics of the devices.

### A. Strain relaxation

The InAs channel QWFETs are incorporated into an $In_{0.53}Ga_{0.47}As/In_{0.52}Al_{0.48}As$ heterostructure system as depicted in Fig. 2, which is epitaxially grown on an InP substrate. The $In_{0.53}Ga_{0.47}As$ and $In_{0.52}Al_{0.48}As$ layers are lattice matched to the InP substrate. InAs and InP, however, have a lattice mismatch of 3.2%, which gives rise to a biaxial compressive strain in the InAs channel region. Such biaxial compressive strain is known to increase the band gap of the InAs quantum well [9]. The valence-force-field (VFF) method with a modified Keating potential is used to compute the relaxed atom positions in the strained heterostructure [13, 17, 18]. The dimensions of the strain relaxation domain are given in Fig. 2(a). Since strain is a long-range effect, a 40 nm thick InAlAs layer below the InGaAs sub-channel is included in the strain relaxation domain. Periodic boundary conditions are applied to the axes perpendicular to the growth direction. Their dimensions are $L_x = L_z = 3.5$ nm. This domain contains 30,816 atoms, corresponding to 3,852 zincblende unit cells and it is sufficiently large to model the random placement of the cations in the InGaAs and InAlAs layers [14] and to extract a transport and confinement effective mass, as shown later.

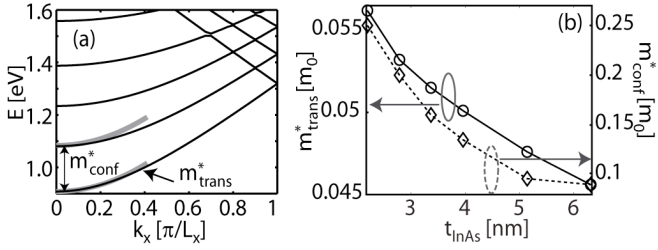

Fig. 3. Bandstructure and effective mass of the InAs quantum well. (a) Bandstructure (at $k_z$=0) of a 5 nm thick InAs quantum well embedded between InGaAs and InAlAs barriers. (b) Transport ($m^*_{trans}$) and confinement ($m^*_{conf}$) effective masses as function of the InAs quantum well thickness. The effective masses are higher in thinner InAs quantum wells due to stronger quantization and a high degree of InAs non-parabolicity.

In Fig. 2(b), the lattice constants of the relaxed heterostructure are compared to the unstrained InAs lattice constant. The in-plane lattice constant along the center of the InAs quantum well ($a_x$,$a_z$) is compressed to the lattice constant of the InP substrate, causing an in-plane biaxial compressive strain of $\varepsilon_\perp = a_x/a_{InAs} - 1 = -0.031$. The lattice constant along the growth direction ($a_y$) is extended to 0.6207 nm in the InAs quantum well region, which corresponds to an orthogonal tensile strain $\varepsilon_\parallel = 0.025$. The value of InAs Poisson ratio in this heterostructure simulation is $\nu=\varepsilon_\parallel/\varepsilon_\perp$=0.806, which is slightly smaller than the bulk value of 1.088 [19]. The ($a_y$) fluctuations in the InGaAs and InAlAs regions are induced by the local bond length variation due to the random placement of the In, Ga, and Al cations and by the bi-modal In-As and Ga-As/Al-As bond length distribution shown in Fig. 2(c). The bimodal In-As and Ga-As bond distribution is a well-known effect and is critical to accurately model strain distribution and its effects on electron transport and optical spectra in InGaAs heterostructures [20] [21]. InAlAs exhibits a similar bi-modal distribution as shown in Fig. 2(c). The inhomogeneous strain fields resulting from the bimodal distribution are included in the electronic structure calculation to correctly model the wavefunction penetration into the InGaAs and InAlAs barriers.

### B. Tight-binding based channel effective mass extraction

An accurate computation of the channel effective mass is critical in devices subject to strain and strong bandstructure non-parabolicities, as is the case of InAs. The effective mass determines the channel properties such as injection velocity, source-to-drain tunneling, quantum capacitance, and density-of-states [22]. Here, the effective masses of the multi-quantum-well channel are extracted from a $sp^3d^5s^*$ tight-binding bandstructure that includes strain. The general purpose electronic structure simulator NEMO-3D [12-14] is used for this computation. The InAs, GaAs, and AlAs tight-binding parameters are taken from Refs. [12, 23]. The bulk parameters are fully transferable to nanostructures and have previously been benchmarked against complex experimental devices such as InAs/InGaAs/InAlAs quantum dots [21] and InAs QWFETs [9].

The tight-binding electronic structure calculation domain (dotted rectangle in Fig. 2(a)) is smaller than the strain relaxation domain. Only 2 nm thick portions of the InAlAs layers on the top and the bottom of the In$_{0.53}$Ga$_{0.47}$As(2 nm)/InAs/In$_{0.53}$Ga$_{0.47}$As(3 nm) quantum-well channel are included since the penetration of the wavefunction beyond this domain is negligible. The in-plane dimensions are the same as the strain relaxation domain. The electronic structure domain contains 7,776 atoms. The bandstructure of the quantum-well active region with a 5 nm thick InAs layer is shown in Fig. 3(a). The effect of strain and quantization due to band discontinuities at the InAs/InGaAs and InGaAs/InAlAs heterostructure interfaces are automatically included due to the atomistic nature of the tight-binding Hamiltonian. The bands shown in Fig. 3(a) are the $\Gamma$ valley sub-bands. The $L$ valley sub-bands (not shown) are at least 0.9 eV higher than the lowest $\Gamma$ valley sub-band for the InAs channel thicknesses considered here, which range from 2.2 nm to 6.2 nm. Due to the large energy separation and the low supply voltages that are applied to the devices ($V_{DD}$=0.5 V), the $L$ valleys do not affect the operation of InGaAs based QWFETs [10] and are safely ignored in the transport calculations.

The $\Gamma$ valley transport effective mass ($m^*_{trans}$) is extracted by fitting a parabola to the lowest conduction sub-band while the confinement effective mas ($m^*_{conf}$) is fitted to replicate the energy difference between the first two tight-binding sub-bands in the effective mass calculation (Fig. 3(a)). In the effective mass calculation the domain of the same size as the tight-binding electronic structure domain is used. The InGaAs and InAlAs layers around the InAs channel are included in the effective mass calculation since the band discontinuities at the interfaces and the wavefunction penetration into these layers affect the confinement in the channel. The values of the band-offsets at the heterostructure interfaces in the effective mass calculation are $\Delta E_{C,InGaAs/InAs} = 0.4$ eV and $\Delta E_{C,InGaAs/InAlAs} = 0.5$ eV, while the transport effective masses of In$_{0.53}$Ga$_{0.47}$As and In$_{0.52}$Al$_{0.48}$As are 0.041·$m_0$ and 0.075·$m_0$ respectively [24, 25]. The In$_{0.53}$Ga$_{0.47}$As and In$_{0.52}$Al$_{0.48}$As effective masses along the growth direction are used as fitting parameters to obtain the correct gate leakage current as explained later.

Such an elaborate, atomistic based fitting procedure allows for an accurate determination of the channel transport and confinement effective masses. As shown in Fig. 3(b), both the transport and confinement effective masses in the InAs quantum well are significantly larger than their bulk value ($m^*_{InAs} = 0.023·m_0$) due to the strong quantum confinement in the strongly non-parabolic InAs/InGaAs system, the wavefunction penetration into the heavier effective mass InGaAs and InAlAs barrier layers, and the biaxial strain. The effective masses become heavier as the quantum well thickness is reduced emphasizing the strong non-parabolic dispersion of InAs [26].

### C. 2-D Real Space Effective Mass Simulator

The Schrödinger and Poisson equations are solved self-consistently using the effective mass approximation on a 2-D finite-difference grid [27]. The grid is uniform and the spacing along the $x$ and $y$-directions are $\Delta x = 0.25$ nm and $\Delta y = 0.2$ nm, respectively. The quantum transport simulation domain is shown in Fig. 1. In the Poisson calculation, Von Neumann boundary conditions are applied everywhere except at the gate contact, where Dirichlet boundary conditions are applied. In the ballistic transport model used here, electrons are injected into the device from the source, drain, and gate contacts at different energy values and the resulting contributions are summed up to obtain the carrier and current densities. The real space technique accurately accounts for the longitudinal ($x$-axis) and transverse ($y$-axis) mode coupling [27] and for gate leakage currents.

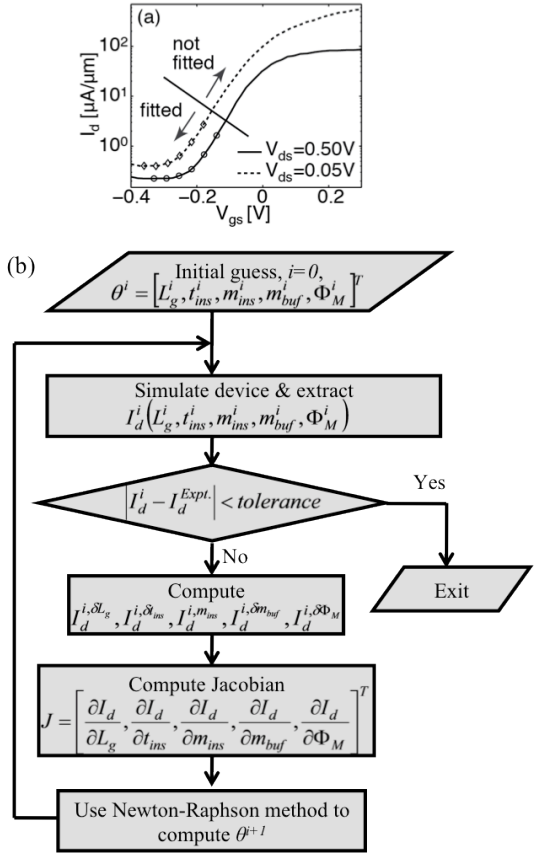

Fig. 4. Parameter fitting procedure. (a) Experimental $I_d$-$V_{gs}$ characteristics of the 30 nm gate length device. The leakage and the subthreshold regimes identified by the markers are used in the fitting procedure. (b) Flow chart of the parameter fitting procedure.

### D. Calibration methodology

To calibrate the simulator to the experimental data, five fitting parameters are used: (i) $L_g$ – the gate length, (ii) $t_{ins}$ – the thickness of the InAlAs insulator layer between the quantum-well channel and the gate contact, (iii) $m^*_{ins}$ – the effective mass of the InAlAs insulator along the growth direction ($y$), (iv) $m^*_{buf}$ – the effective mass of the InGaAs sub-channels along the growth direction ($y$), and (v) $\Phi_M$ – the gate metal work function. The leakage and sub-threshold (low $V_{gs}$) regimes of the $I_d$-$V_{gs}$ characteristics are chosen as fitting regions because the currents are very sensitive to the device dimensions and material parameters there, but they do not depend on the source and drain series resistances. As a result, the gate leakage current and the subthreshold slope are properly modeled. The sensitivity of the device performance to the fitting parameters is discussed in the appendix. The currents at high $V_{gs}$ are mainly governed by the source and drain series resistances, which are not adjusted but set to the experimentally measured values $R_S = 0.21$ $\Omega$·mm and $R_D = 0.23$ $\Omega$·mm.

The selection of the fitting parameters is based on two criteria: (i) the fabrication process variability and (ii) the sensitivity to the drain current. The gate length $L_g$ and insulator thickness $t_{ins}$ are respectively determined by lithography and wet chemical etching processes which are prone to variability [7, 28]. Likewise, the difficulty of controlling the surface conditions before metal deposition induces process variability in the gate metal work function $\Phi_M$ [6, 29]. The effective masses of the In$_{0.53}$Ga$_{0.47}$As and In$_{0.52}$Al$_{0.48}$As top layers along the growth direction ($m^*_{buf}$ and $m^*_{ins}$ respectively) are not accurately known from the experimental measurements and are also included in the fitting procedure. Their values, however, are allowed to vary only within the experimentally reported ranges [24, 25]. The 2-D electrostatics is governed by $L_g$ and $t_{ins}$, while $t_{ins}$, $m^*_{buf}$, $m^*_{ins}$, and $\Phi_M$ control the electron tunneling probability from the gate into the InAs channel, which in turn determines the gate leakage current.

The thickness of the InAs channel and the InGaAs buffer are not included in the fitting procedure because they are determined by the Molecular Beam Epitaxy (MBE) growth, which is a precise atomic layer deposition technique. The electron affinity of the InAs channel ($\chi$ = 4.9 eV [25]) and the conduction band offsets at the heterostructure interfaces ($\Delta E_{C,InGaAs/InAs}$ = 0.4 eV and $\Delta E_{C,InGaAs/InAlAs}$ = 0.5 eV [24, 25]) are not considered as fitting parameters because the drain current is less sensitive to them.

The fitting procedure is summarized in Fig. 4. The drain current is parameterized as $I_d(L_g, t_{ins}, m^*_{ins}, m^*_{buf}, \Phi_M)$, were, $L_g$ is the gate length, $t_{ins}$ is the thickness of the InAlAs layer between the quantum-well channel and the metal gate, $m^*_{ins}$ is the effective mass of InAlAs, $m^*_{buf}$ is the effective mass of the InGaAs buffer, and $\Phi_M$ is the gate metal work function. An iterative approach is used where, at each iteration, the subthreshold $I_d$-$V_{gs}$ current characteristics of the considered devices are computed and compared to the experimental data. If the deviation from the experimental data is larger than the tolerance, a new guess to the parameter vector is computed using a Newton-Raphson scheme [30]. Since an analytical expression for the drain current in terms of the fitting parameters does not exist, the partial derivatives composing the Jacobian matrix are computed numerically. To evaluate a partial derivative with respect to a parameter X, the subthreshold $I_d$-$V_{gs}$ characteristics of a device whose parameter X has been increased by $\Delta$X are simulated. All the other parameters remain the same as in the reference device. For example, the partial derivative with respect to the gate length

$L_g$ is $\dfrac{\partial I_d}{\partial L_g} = \dfrac{I_d^{i,\delta L_g} - I_d^i}{\delta L_g}$ where, $i$ – the iteration count, $I_d^i$ – the drain current of the reference device parameterized as $I_d^i(L_g^i, t_{ins}^i, m_{ins}^{*,i}, m_{buf}^{*,i}, \Phi_M^i)$, $\delta L_g$ – change in the gate length of a new device from the reference device, and $I_d^{i\delta L_g}$ – the drain current of a new device parameterized as $I_d^{i\delta L_g}(L_g^i + \delta L_g, t_{ins}^i, m_{ins}^{*,i}, m_{buf}^{*,i}, \Phi_M^i)$. The same procedure is repeated to compute the partial derivatives with respect to $t_{ins}$, $m^*_{ins}$, $m^*_{buf}$, and $\Phi_M$. The parameter shifts used to compute the numerical derivatives are ($\delta L_g$, $\delta t_{ins}$, $\delta m^*_{ins}$, $\delta m^*_{buf}$, $\delta\Phi_M$) = (0.5 nm, 0.2 nm, 0.005·$m_0$, 0.005·$m_0$, 0.05 eV), where $m_0$ is the free electron mass. For each device, the voltage sweep shown in Fig. 4(a) requires typically 4 hours on 40 cores on a 2.5 GHz quad core AMD 2380 processor [31].

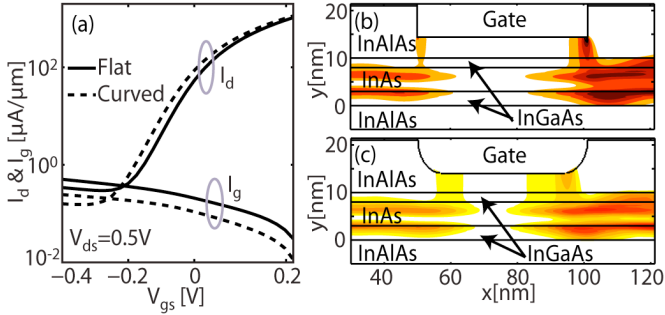

Fig. 5. Influence of the gate geometry on the gate leakage. (a) Intrinsic $I_d$-$V_{gs}$ and $I_g$-$V_{gs}$ characteristics of an InAs QWFET ($L_g$ = 51nm) with a flat (solid lines) and a curved (dashed lines) gate contact. (b) OFF-state current distribution in a flat gate contact device. (c) OFF-state current distribution in a curved gate contact device. The same bias conditions ($V_{gs}$ = -0.4 V and $V_{ds}$ = 0.5 V) and the same color scheme is used in (b) and (c). The magnitude of the current decreases from dark to light colors.

## IV. RESULTS

The methodology presented in Section III is first used to calibrate the simulator against experimental $I_d$-$V_{gs}$ from devices with gate lengths ranging from 30 nm to 50 nm [7]. The calibration phase can be seen as a metrology experiment, where the actual gate lengths and material properties are estimated. The resulting calibrated simulator is then used to optimize the design of a 20 nm gate length device.

### A. Comparison to experimental data

Gate leakage currents are much larger in QWFETs than in conventional Si MOSFETs due to the absence of a proper insulator layer such as $SiO_2$ or $HfO_2$. Moreover, the gate contact geometry plays an important role in determining the magnitude of the gate leakage currents. The shape of the gate contact depends on the fabrication technique used to thin down the insulator before deposing the gate metal stack. Anisotropic etching and metal gate sinking ideally lead to a flat gate contact while isotropic etching leads to a curved gate contact (Fig. 1).

Flat or curved gate contact geometries act differently on the leakage current magnitude as illustrated in Fig. 5 (a). Both devices perform similarly in the high $V_{gs}$ regime, however, their currents significantly differ at low $V_{gs}$. The flat gate device exhibits a lower subthreshold slope $SS$ = 83.5 mV/dec as compared to the curved gate device ($SS$ = 89.7 mV/dec), but its gate leakage current is about 2 times higher. The

| Parameter | Initial | Final parameter set | | |
|---|---|---|---|---|
| | | 30 nm | 40 nm | 50 nm |
| $L_g$ [nm] | 30, 40, 50 | 34.0 | 42 | 51.25 |
| $t_{ins}$ [nm] | 4 | 3.6 | 3.8 | 4.0 |
| $m^*_{buf}$ [$m_0$] | 0.075 | 0.078 | 0.078 | 0.078 |
| $m^*_{ins}$ [$m_0$] | 0.041 | 0.043 | 0.043 | 0.043 |
| $\Phi_M$ [eV] | 4.7 | 4.660 | 4.693 | 4.678 |

Typically convergence is achieved in less than 15 iterations. Here, $m_0$ is the free electron effective mass.

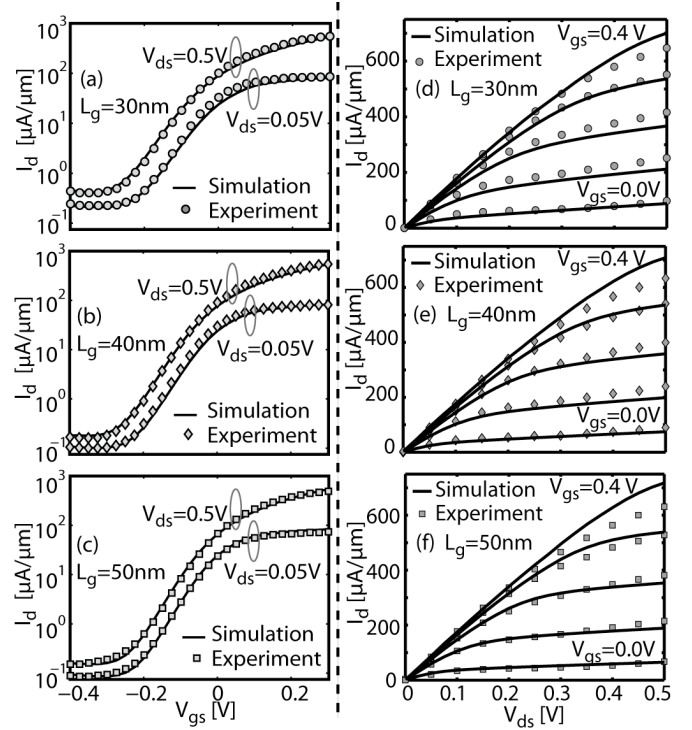

Fig. 6. Comparison between the experimental and simulated $I_d$-$V_{gs}$ characteristics of (a) 30 nm, (b) 40 nm, and (c) 50 nm, $I_d$-$V_{ds}$ characteristics of (d) 30 nm, (e) 40 nm, and (f) 50 nm gate length InAs QWFETs. The $I_d$-$V_{ds}$ characteristics in figures (d)-(f) are calculated at $V_{gs}$ of 0.0 V, 0.1 V, 0.2 V, 0.3 V, and 0.4 V, where only 0.0 V and 0.4 V are labeled. The device dimensions and material parameters in Table I are used in the simulations.

difference between the drain currents at low $V_{gs}$ is the result of gate leakage ($I_g$) suppression in the curved gate contact device.

The current distribution in the gate leakage regime is shown in Fig. 5(b,c). Gate leakage currents are concentrated at the edges of the contact due to lower tunneling barriers and higher electric fields there as compared to the central region of the gate contact [11]. The curved gate device is characterized by a thicker insulator at the edges of the gate contact, which leads to a suppression of the leakage current. Thus, an accurate description of the gate contact geometry is crucial to reproduce the experimental $I_d$-$V_{gs}$, especially in the leakage and subthreshold regimes, and can be seen as a design parameter. A curved gate geometry is clearly seen in the TEM micrographs of Ref. [7]. A curved gate geometry, which resembles that of the experimental devices, is therefore used in the benchmarking procedure. As shown in Fig. 1, the shape of the edges of the curved gate contact is a quarter circle with the

| $L_g$ [nm] | | $I_{OFF}$ [μA/μm] | $SS$ [mV/dec] | $DIBL$ [mV/V] | $I_{ON}/I_{OFF}$ | $v_{inj}$ [cm/s] |
|---|---|---|---|---|---|---|
| 30 | Expt. | 0.4077 | 106.9 | 168.9 | $0.47×10^3$ | |
| | Sim. | 0.4431 | 105.2 | 144.7 | $0.61×10^3$ | $3.0×10^7$ |
| 40 | Expt. | 0.1607 | 90.9 | 126.0 | $1.38×10^3$ | |
| | Sim. | 0.1789 | 89.4 | 99.3 | $1.86×10^3$ | $3.11×10^7$ |
| 50 | Expt. | 0.1696 | 85.1 | 97.2 | $1.80×10^3$ | |
| | Sim. | 0.1519 | 89.2 | 90.8 | $1.85×10^3$ | $3.18×10^7$ |

The procedure of Ref. [32] is used to extract the device performance parameters. The threshold voltage ($V_T$) is defined as the $V_{gs}$ yielding $I_d = 1$ μA/μm.

curvature radius equal to $t_{InAlAs}$-$t_{ins}$, which changes as $t_{ins}$ changes in the fitting procedure.

The results of the device metrology for devices with gate lengths ranging from 30 nm to 50 nm are summarized in Table I. Here, the transport and confinement effective mass values extracted from the tight-binding bandstructures are used in the InAs channel region, which for a 5 nm thick InAs channel amount to $m*_{trans} = 0.049 \cdot m_0$ and $m*_{conf} = 0.096 \cdot m_0$ respectively (Fig. 3(b)). The parameter values at the end of the fitting procedure are close to the experimentally reported values, which are used as an initial guess [7]. The effective masses of the InGaAs buffer ($m*_{buf}$) and the InAlAs insulator ($m*_{ins}$) layers are not allowed to vary between the different devices since they are all fabricated side by side on the same heterostructure. The effective mass values after convergence of the fitting process are within the ranges reported in the literature, which are $0.038 \cdot m_0 - 0.044 \cdot m_0$ for $In_{0.53}Ga_{0.47}As$ and $0.070 \cdot m_0 - 0.083 \cdot m_0$ for $In_{0.52}Al_{0.48}As$ [24, 25]. The converged values for the gate length and insulator thickness are within the expected process variability of the wet chemical etching step used to thin down the InAlAs insulator prior to gate metal deposition [7]. The InAlAs insulator thickness and metal work function values automatically adjust to match the different magnitude of the gate leakage current in each device (Fig. 6). Slightly different values of the gate metal work function are justified because of the lack of precise control of the surface oxides, which modify the work function value [7].

The experimental transfer $I_d$-$V_{gs}$ and output $I_d$-$V_{ds}$ characteristics are compared to the simulation results in Fig. 6. The parameters given in Table I are used in these simulations. It should be noted that only the low $V_{gs}$ regime is used in the fitting procedure as described in Fig. 4. The performance parameters extracted from the experimental and simulated $I_d$-$V_{gs}$ in Fig. 6 agree reasonably well, as shown in Table II. The injection velocity ($v_{inj}=I_{ON}/Q_{top}$) is calculated at the top of the potential barrier in the InAs channel [22].

At high biases, the current-voltage characteristics of the QWFETs are dominated by the source and drain contact resistances which are modeled as two external series resistances $R_S$ and $R_D$ attached to the intrinsic device (Fig. 1) following the procedure described in Ref. [15]. The $I_d$-$V_{gs}$ and $I_d$-$V_{ds}$ characteristics of the intrinsic device (Fig. 1) are calculated for the bias ranges $-0.4 \leq V_{gs}^{int} \leq 0.3$ V and $0 \leq V_{ds}^{int} \leq 0.5$ V. The extrinsic device characteristics are then computed from the drain current, $I_d$, at given extrinsic terminal voltages ($V_{gs}$ and $V_{ds}$) using $V_{gs} = V_{gs}^{int} + R_S I_d$ and

$V_{ds} = V_{ds}^{int} + (R_S + R_D)I_d$. The experimentally reported values of $R_S = 0.21$ Ω·m and $R_D = 0.23$ Ω·m are used [7]. The good quantitative agreement shown in Fig. 6 is enabled by the consideration of a curved gate contact geometry, an accurate estimation of the channel effective masses, as well as the parameter adjustments listed in Table I.

The simulated $I_d$-$V_{ds}$ characteristics show a very good agreement with the experimental $I_d$-$V_{ds}$ at low bias voltages while $I_d$ is overestimated at high bias voltages (Fig. 6). The overestimation of $I_d$ is related to the fact that scattering is not included in the ballistic quantum transport model. Electron-phonon, interface roughness, and alloy disorder scattering appear to play a non-negligible role at high biases. The deviation between the simulated and experimental $I_d$ is larger in devices with a longer gate contact. In effect, the electron transport in longer devices is more affected by scattering than in shorter devices, which operate closer to their ballistic limit.

### B. Design optimization of a 20 nm device

After benchmarking the simulation approach and providing metrology insight into experimental devices, we will explore the performance of a hypothetical 20 nm gate length device. The effects of the InAs channel thickness ($t_{InAs}$), the InAlAs insulator thickness ($t_{ins}$), and the gate metal work function ($\Phi_M$) are investigated. The scaling of $t_{InAs}$ and $t_{ins}$ improves the gate control of the channel surface potential while a higher $\Phi_M$ suppresses the gate leakage and shifts the threshold voltage ($V_T$) in positive direction towards enhancement mode operation.

A flat gate contact geometry provides a superior gate control of the channel potential as compared to a curved contact (Fig. 5) at the price of higher gate leakage current. Since the leakage can be reduced through work function engineering, as explained later, a flat gate geometry will be used in the performace evaluation of the 20 nm gate length device. A flat gate contact can be "easily" fabricated by replacing the isotropic wet chemical etching step used to thin down the InAlAs insulator layer by an anisotropic etching or by a metal gate sinking technique [6]. These advanced fabrication techniques will result in smaller radius of curvature or near ideal flat contact.

The performance parameters ($SS$, $DIBL$, and $I_{ON}/I_{OFF}$ ratio) are calculated by using the constant overdrive voltage method proposed in Ref. [32]. The threshold voltage $V_T$ is determined from a linear extrapolation of the $I_d$-$V_{gs}$ characteristics at the peak transconductance to zero $I_d$ (maximum-$g_m$ method [33]) and a supply voltage $V_{DD}$ of 0.5 V is used. The ON-state is defined as $V_g = V_T + 2V_{DD}/3$, $V_d = V_{DD}$, $V_s = 0$ while the OFF-state is defined as $V_g = V_T - V_{DD}/3$, $V_d = V_{DD}$, $V_s = 0$. The capacitances are defined as (i) the gate capacitance: $C_g = dQ_s/dV_{gs}$, (ii) the insulator capacitance: $C_{ins} = \varepsilon_{ins}/t_{ins}$, and (iii) the inversion layer capacitance: $C_{inv} = dQ_s/d\psi_s$ [34]. Here, $Q_s$ is the sheet charge density in the InGaAs/InAs/InGaAs composite channel, $\varepsilon_{ins}$ the dielectric constant of the InAlAs insulator, and $\psi_s$ the surface potential at the interface between the InAlAs insulator and the InGaAs/InAs/InGaAs composite

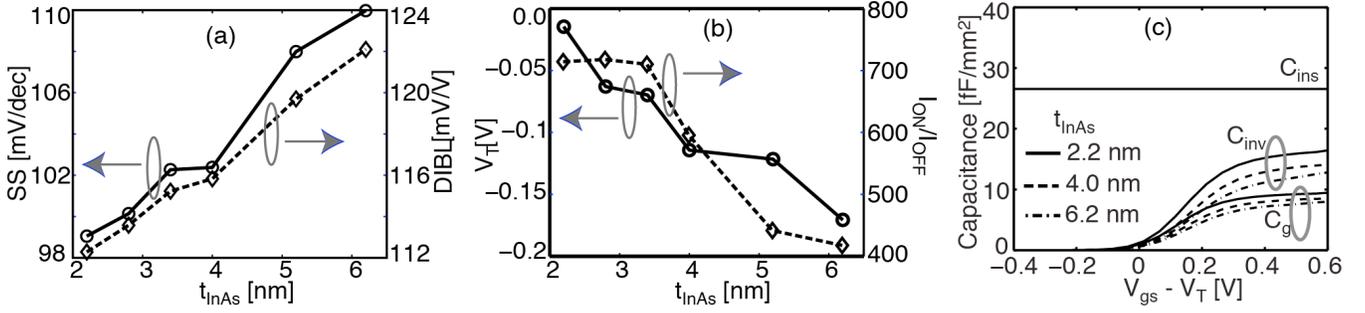

Fig. 7. InAs quantum well thickness ($t_{InAs}$) scaling. (a) *SS* and *DIBL*. (b) $V_T$, and $I_{ON}/I_{OFF}$ ratio. (c) Gate capacitance ($C_g$) and inversion layer capacitance ($C_{inv}$) as function of the gate overdrive ($V_{gs}$-$V_T$) for devices with a fixed $t_{ins}$ = 4 nm and a variable $t_{InAs}$. The device dimensions, except $t_{InAs}$, are the same as in Fig. 1 with $L_g$ = 20 nm and $t_{ins}$ = 4 nm. The gate metal work function is set to $\Phi_M$ = 4.7 eV.

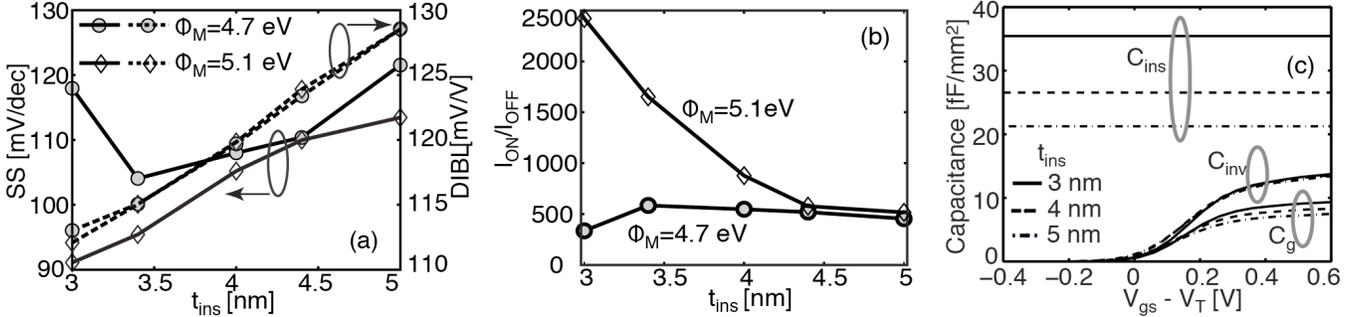

Fig. 8. InAlAs insulator thickness ($t_{ins}$) scaling. (a) *SS* and *DIBL* of devices with $\Phi_M$ = 4.7 eV and 5.1 eV. (b) $I_{ON}/I_{OFF}$ ratio of the same devices. (c) Gate capacitance ($C_g$) and inversion layer capacitance ($C_{inv}$) as function of the gate overdrive ($V_{gs}$-$V_T$) for devices with a fixed $t_{InAs}$ = 5 nm and a variable $t_{ins}$. The same device dimensions (except $t_{ins}$) as in Fig. 1 are used with $L_g$=20 nm.

channel.

### 1) InAs channel thickness ($t_{InAs}$)

The effect of scaling down the InAs channel thickness ($t_{InAs}$) can be analysed by noting that the QWFET is electrostatically very similar to a fully depleted silicon on insulator (FD-SOI) MOSFET [28]. In a QWFET, the InAs channel thickness plays a role similar to the Si body thickness of an FD-SOI MOSFET. Higher gate length to channel thickness ratio in a thin InAs channel QWFET results in a stronger gate control of the channel surface potential, which improves *SS* and *DIBL* (Fig. 6(a)). This is similar to ultra-thin Si body FD-SOI MOSFETs [35].

The strong electrostatic confinement of electrons in thin InAs quantum well devices pushes the channel conduction subbands to higher energies, which subsequently results in higher $V_T$ and facilitates enhancement mode operation of such devices (Fig. 6(b)). A similar $V_T$ shift caused by scaling down the Si body thickness is observed in FD-SOI MOSFETs [36].

The channel effective mass along the transport direction increases as $t_{InAs}$ decreases (Fig. 3), which leads to a lower electron injection velocity ($v_{inj}$), but a higher carrier density ($N_{inj}$) at the QWFET virtual source [37]. The net effect is a higher $I_{ON}$ in thin InAs channel devices. The 2D electron gas in thinner channel devices is located closer to the gate, resulting into a higher gate leakage and $I_{OFF}$. As $t_{InAs}$ is slightly reduced, the higher $I_{ON}$ more than compensates the larger $I_{OFF}$, so that the $I_{ON}/I_{OFF}$ ratio actually increases. However, if $t_{InAs}$ is further reduced, the increase of the OFF current becomes more important and the $I_{ON}/I_{OFF}$ ratio starts to saturate (Fig. 6(b)).

The effect of $t_{InAs}$ on the total gate capacitance, $C_g$ is depicted in Fig. 7(c). It can be observed that $C_g$, which is the series combination of the insulator $C_{ins}$ and inversion $C_{inv}$ capacitances, increases as $t_{InAs}$ decreases due to the increase of the channel effective mass. Hence, the inversion charge $N_{inv}$ at the virtual source is larger in thin channel devices, $N_{inv}$ being directly proportional to $C_g$. The increase in $C_{inv}$ in thin $t_{InAs}$ devices is attributed to increase in both of its components, namely the quantum capacitance $C_Q$ and the centroid capacitance $C_{cent}$. $C_Q$ increases because of higher effective mass while $C_{cent}$ increases because the electron gas is closer to the gate in thin $t_{InAs}$ devices [26].

### 2) InAlAs insulator thickness ($t_{InAlAs}$)

The scaling of the InAlAs insulator thickness ($t_{ins}$) in devices with $\Phi_M$ of 4.7 eV and 5.1 eV is illustrated in Fig. 8. When $\Phi_M$ = 4.7 eV, the performance metrics *SS*, *DIBL*, and $I_{ON}/I_{OFF}$ ratio improve as $t_{ins}$ is scaled down until 3.4 nm. This can be attributed to a better electrostatic control of thin insulator devices. When $t_{ins}$ is scaled below 3.4 nm, *DIBL* keeps decreasing, while the *SS* and $I_{ON}/I_{OFF}$ ratio, which are affected by the electron tunneling across the InAlAs insulator, start degrading due to an excessive gate leakage. This degradation can be controlled by increasing $\Phi_M$ to 5.1 eV, which increases the tunneling barrier height between the gate and the InAs channel, reduces the gate leakage, and therefore improves the *SS* and $I_{ON}/I_{OFF}$ ratio even with the InAlAs layer scaled down to 3 nm (Fig. 8(a,b)). The mechanism of the gate leakage suppression through metal work function engineering is discussed in the next sub-section.

Since III-V devices are characterized by a small density-of-states effective mass and therefore a small inversion capacitance $C_{inv}$, the down scaling of $t_{ins}$ does not significantly increase the total gate capacitance $C_g$ as shown in Fig. 8(c). Similar trends in $C_g$ are observed in the experiments reported in Ref. [26].

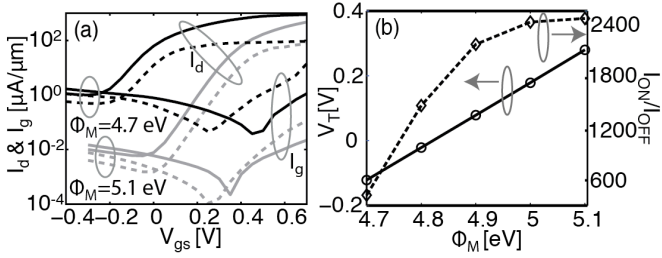

Fig. 9. Gate leakage reduction and enhancement-mode operation by gate metal work function ($\Phi_M$) engineering. (a) $I_d$-$V_{gs}$ and $I_g$-$V_{gs}$ characteristics of $L_g$ = 20 nm InAs QWFETs with $\Phi_M$ = 4.7 eV and 5.1 eV. (b) $V_T$ and $I_{ON}/I_{OFF}$ ratio as function of $\Phi_M$ for a $L_g$ = 20 nm InAs QWFET. The device dimensions are the same as in Fig. 1 with $L_g$ = 20 nm and $t_{ins}$ = 3 nm.

### 3) Gate metal work function ($\Phi_M$)

As shown in Fig. 8(a,b), a device with $\Phi_M$ = 5.1 eV shows a significant improvement in $SS$ and $I_{ON}/I_{OFF}$ ratio compared to a device with $\Phi_M$ = 4.7 eV. To explain this effect the $I_d$-$V_{gs}$ and $I_g$-$V_{gs}$ characteristics of devices with the same $t_{ins}$ = 3 nm, but different $\Phi_M$ (4.7 and 5.1 eV) are compared in Fig. 9. The longitudinal (x-axis) and transverse (y-axis) band-diagrams of the same devices in the OFF-state ($V_{ds}$ = $V_{DD}$, $V_{gs}$ - $V_T$ = - $V_{DD}/3$) are shown in Fig. 10. Under the same bias conditions, the device with $\Phi_M$ = 5.1 eV shows a 100× smaller gate leakage current ($I_g$) as compared to the device with $\Phi_M$ = 4.7 eV. The mechanism of the gate leakage suppression in high $\Phi_M$ device is explained in Fig. 10. In the transistor OFF-state, the gate electrons see a different tunneling barrier depending on the gate metal work function. With $\Phi_M$ = 4.7 eV, the gate electrons must tunnel through the InAlAs insulator layer only to reach the InAs channel while they must tunnel through the InAlAs and InGaAs layers when $\Phi_M$ = 5.1 eV, considerably reducing the gate leakage current.

Although $I_g$ of the device with $\Phi_M$ = 5.1 eV shows a 100× reduction, its $I_{ON}/I_{OFF}$ ratio shows only a 7× improvement compared to the device with $\Phi_M$ = 4.7 eV (Fig. 8(b)). This is due to the fact that the OFF-current, when $\Phi_M$ = 5.1 eV, is no longer dominated by gate leakage currents, but by the thermionic emission of electrons from the source to the drain.

In addition to a larger $I_{ON}/I_{OFF}$ ratio, a higher $\Phi_M$ pushes the threshold voltage $V_T$ towards positive values. In Fig. 9, the $V_T$ values of the devices with $\Phi_M$ = 4.7 eV and $\Phi_M$ = 5.1 eV are - 0.11 V and 0.29 V respectively. The positive shift in $V_T$ is equal to the work function difference $\Delta\Phi_M$ = 0.4 eV. Such positive $V_T$ shift is highly desirable for the enhancement mode operation of the n-type FET in CMOS logic applications [4, 6]. The variation of the $I_{ON}/I_{OFF}$ ratio and $V_T$ for the intermediate values of $\Phi_M$ are shown in Fig 9(b). A higher $\Phi_M$ linearly pushes $V_T$ towards a positive value while the $I_{ON}/I_{OFF}$ ratio increases in a non-linear fashion. Higher $\Phi_M$ values can be achieved by using the Platinum (Pt) buried-gate technology [6, 29].

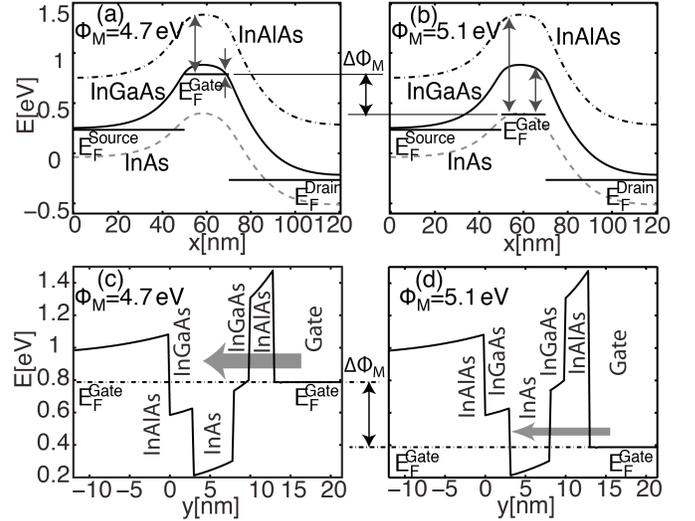

Fig. 10. Gate leakage reduction through gate metal work function ($\Phi_M$) engineering. (a) Conduction band diagrams along three horizontal lines through (i) the center of the InAlAs insulator (dashed-dotted line), (ii) the top InGaAs barrier (solid line), and (iii) the InAs channel (dashed line) of a $L_g$ = 20 nm InAs QWFET with $\Phi_M$ = 4.7 eV at $V_{gs}$ - $V_T$ = -$V_{DD}/3$ = -0.1667 V and $V_{ds}$ = 0.5 V. (b) Same as (a) but for $\Phi_M$ = 5.1 eV. The same source/drain Fermi levels and band bending, but different gate voltages, are used in (a) and (b), ensuring almost the same source to drain current in both devices. Electrons tunneling from the gate terminal into the channel experience a higher energy barrier (arrows) in the $\Phi_M$ = 5.1 eV device as compared to the $\Phi_M$ = 4.7 eV device because of the metal gate Fermi level offset equal to the work function difference ($\Delta\Phi_M$). (c) Band diagram of the device in (a) along a vertical line near the drain side edge of the gate contact. (d) Same as (c) but for $\Phi_M$ = 5.1 eV. The electrons in the $\Phi_M$ = 4.7 eV device tunnel only through the InAlAs insulator while the electrons in the $\Phi_M$ = 5.1 eV device must tunnel through the InGaAs buffer layer in addition to the InAlAs insulator. The thickness of the arrows in (c,d) schematically shows the direction and the magnitude of the electron tunneling current.

## V. CONCLUSION

Performances of InAs QWFETs have been analyzed by using a multiscale device simulation approach. The effective mass of the InAs channel raises significantly from its bulk value due to strong confinement effects, which are included on the atomistic scale through a $sp^3d^5s*$ tight-binding model. The gate tunneling is critical in the device analysis and is included in a 2-D Schrödinger-Poisson solver by injecting carriers from the gate contact in addition to the source and drain contacts. The simulation approach is calibrated against experimental devices with gate lengths ranging from 30 to 50 nm. A good quantitative match between the experimental and simulated current-voltage characteristics is reported. The accurate description of the shape of the gate contact is essential to replicate the experimental results.

The calibrated simulation methodology has been used to investigate the design optimizations of a hypothetical 20 nm gate length InAs QWFET. The scaling of the InAs channel thickness and the InAlAs insulator thickness improve the logic performance due to a stronger gate control of the channel potential. An excessive scaling, however, leads to higher gate leakage current, which degrades the device performances. The

| Device | | $L_g$ [nm] | $t_{ins}$ [nm] | $m^*_{buf}$ [m0] | $m^*_{ins}$ [m0] | $\Phi_M$ [eV] |
|---|---|---|---|---|---|---|
| 30 nm | $I_{OFF}$ | -0.01 | 2.91 | 5.88 | 0.83 | 35.45 |
| | SS | 1.03 | 1.33 | 1.71 | 1.52 | 11.36 |
| 40 nm | $I_{OFF}$ | -0.19 | 6.17 | 3.88 | 2.31 | 35.55 |
| | SS | 1.70 | 0.88 | 1.98 | 1.87 | 4.62 |
| 50 nm | $I_{OFF}$ | -0.05 | 6.04 | 3.45 | 1.28 | 36.78 |
| | SS | 2.86 | 1.68 | 3.59 | 3.42 | 5.83 |

The sensitivity of $I_{OFF}$ to $t_{ins}$ is defined as the percentage reduction in $I_{OFF}$ induced by a 1 % increase of $t_{ins}$, while the values of all the other parameters are fixed. The sensitivity of all the other parameters is defined similarly. The parameter sensitivity in each device is calculated with respect to the optimized parameter set given in Table I.

gate leakage current can be suppressed by increasing the gate metal work function, which also pushes the threshold voltage towards the enhancement mode operation. As a result of the reduced gate leakage, high gate metal work function devices can be scaled more aggressively compared to low gate metal work function devices.

The simulation tool, OMEN_FET, that generated the results presented in this paper is available on nanoHUB.org [38].

## APPENDIX

A sensitivity analysis of the fitting parameters is reported in Table III. The sensitivity of each parameter is calculated with respect to the optimized parameter sets given in Table I. The sensitivity is defined as the percentage reduction in $I_{OFF}$ and $SS$ for a 1% increase in the value of each parameter. An increase in the values of $t_{ins}$, $m^*_{ins}$, $m^*_{buf}$, and $\Phi_M$ reduces the tunneling probability from the gate into the channel, leading to a reduction of $I_{OFF}$. An increase of $\Phi_M$ significantly reduces the tunneling probability between the channel and the gate contact, which results in significant reduction of $I_{OFF}$ as well as $SS$. An increase of $L_g$ results in a lower $SS$ as longer gate length devices exhibit a better control of the channel electrostatics. Based on this sensitivity analysis, it can be concluded that transport in the gate leakage and subthreshold regimes is highly dependent on $\Phi_M$, which makes it a critical parameter in the fitting procedure and in the design of ultra-scaled Schottky gated QWFETs.

## ACKNOWLEDGMENT

This work was partially supported by the Semiconductor Research Corporation (SRC) and the MARCO Focus Center on Materials, Structures, and Devices. Computational support was provided by NSF through nanoHUB.org operated by the Network for Computational Nanotechnology (NCN) and National Institute for Computational Sciences (NICS).